\documentclass[twocolumn,floatfix,showpacs,preprintnumbers,amsmath,amssymb,aps]{revtex4}
\usepackage[latin9]{inputenc}
\usepackage{amssymb}
\usepackage{graphicx}
\usepackage{esint}

\makeatletter
\@ifundefined{textcolor}{}
{%
 \definecolor{BLACK}{gray}{0}
 \definecolor{WHITE}{gray}{1}
 \definecolor{RED}{rgb}{1,0,0}
 \definecolor{GREEN}{rgb}{0,1,0}
 \definecolor{BLUE}{rgb}{0,0,1}
 \definecolor{CYAN}{cmyk}{1,0,0,0}
 \definecolor{MAGENTA}{cmyk}{0,1,0,0}
 \definecolor{YELLOW}{cmyk}{0,0,1,0}
}

    \usepackage[english]{babel}

\makeatother

\begin{document}

\title{Entanglement entropy of a scalar field \\
across a spherical boundary in the Einstein universe}

\author{Katja Ried}

\email{kried@perimeterinstitute.ca}

\selectlanguage{english}%

\affiliation{Instituto de F\'{i}sica Te\'{o}rica, Universidade Estadual Paulista, Rua
Dr. Bento Teobaldo Ferraz 271, 01140-070, S\~{a}o Paulo, SP, Brazil}

\affiliation{Perimeter Institute for Theoretical Physics, 31 Caroline St. N. Waterloo
Ontario, N2L 2Y5, Canada}

\date{\today}
\begin{abstract}
A scalar field in the ground state, when partially hidden from observation
by a spherical boundary, acquires entanglement entropy $S$ proportional
to the area of the surface. This area law is well established in flat
space, where it follows almost directly from dimensional arguments.
We study its validity in an Einstein universe, whose curvature provides
an additional physical parameter on which the entropy could, in principle,
depend. The surprisingly simple result is that the entanglement entropy
still scales linearly with the area. This is supported by other observations
to the effect that the entanglement entropy arises mostly from degrees
of freedom near the boundary, making it insensitive to the large-scale
geometry of the background space.
\end{abstract}

\pacs{03.65.Ud, 04.62.+v, 04.70.Dy, 89.70.Cf}

\maketitle

\section{Introduction}

\label{sec:Introduction} 

One of the main open questions in black hole physics concerns the
microscopic explanation (if any) of the black hole entropy law: 
\begin{equation}
S_{{\rm bh}}=A_{{\rm bh}}/(4\, l_{{\rm Pl}}^{2}),\label{eq: S_BH}
\end{equation}
 where $A_{{\rm bh}}$ is the event horizon area and $l_{{\rm Pl}}=\sqrt{G_{{\rm N}}}$
is the Planck length, which equals the square root of the Newtonian
constant $G_{{\rm N}}$ in $c=\hbar=k_{{\rm B}}=1$ units (see Ref.~\cite{W01}
for a comprehensive review and further references). In contrast to
usual thermodynamical systems, the black hole entropy $S_{{\rm bh}}$
scales with the square rather than with the cube of its linear dimension.
Among the various proposals to explain it, the possibility that the
black hole entropy has its roots in the entanglement of the background
fields~\cite{FN93} is perhaps the most conservative one, since it
relies on the same fields which give rise to Hawking radiation. (Nevertheless
some points must still be clarified; see Ref.~\cite{S11} for a recent
review.) This line of thought was raised after Bombelli et al.~\cite{BKLS86}
and Srednicki~\cite{S93} independently concluded that the entanglement
entropy $S$ of a vacuum scalar field across a spherical boundary
in Minkowski space-time scales with its area $A$ rather than with
its volume~\cite{JCP10}. Srednicki also showed that the result is
independent of any infrared cut off, and obtained an explicit result
assuming a massless scalar field: 
\begin{equation}
S\approx0.096\, A/\left(4a^{2}\right),\label{eq: S Srednicki}
\end{equation}
where $a$ is a length quantity derived from the theory's ultraviolet
cut off. 

Some properties of this law can be derived from physical arguments.
For one, the Schmidt decomposition guarantees that the entropy is
the same, regardless of whether we trace over the degrees of freedom
either inside or outside the sphere, provided that the total state
is pure. Thus $S$ can only depend on the area $A$ of the boundary
shared by the two regions, but not, for example, on the radius of
the sphere (which characterizes only the interior) or the infrared
cutoff (exterior). Assuming that the scalar field is massless, the
only other quantity on which the entropy may depend is $a$. By demanding
that $S$ be dimensionless, we constrain the entropy to be a function
of the form $S=S(A/a^{2})$. The fact that, among all possibilities,
Eq.~(\ref{eq: S Srednicki}) turns out to be a linear function of
$A/a^{2}$ is quite remarkable. Here we investigate whether the same
area law~(\ref{eq: S Srednicki}) continues to hold if we introduce
a curved background, which is characterized by an additional parameter
$R_{0}$, the radius of curvature. Then the most general entropy dependence
can be cast as 
\begin{equation}
S=S\left[\frac{A}{4a^{2}},\frac{\pi R_{0}^{2}}{a^{2}}\right].\label{S_general_Einstein}
\end{equation}

The paper is organized as follows. In section~\ref{sec:generalframework},
we specify the problem and outline how the entanglement entropy is
obtained. Sec.~\ref{sec:flat} presents tests of the scheme on regions
with arbitrary geometry in flat (Minkowski) space-time. In Sec.~\ref{sec:Einstein}
we turn to the entanglement entropy of a sphere in the Einstein universe,
and Sec.~\ref{sec:finalremaks} concludes.


\section{General framework}

\label{sec:generalframework} 

Consider, for the sake of simplicity, a ground-state real scalar field
$\varphi\left(\vec{x}\right)$, which lives in a Minkowski space-time.
We compute the entanglement entropy $S$ of a region $\Omega$, which
arises when one traces out the degrees of freedom inside the region
and computes the von Neumann entropy of the reduced state. 

In order to perform this computation numerically, we only consider
a finite number of degrees of freedom, given by the values of the
field at the sites of a finite discrete grid. Let $a$ denote the
spacing of the points, which can be interpreted as an ultraviolet
cutoff: field modes with wavelengths below this limit are not described
by the model. The infrared cutoff, on the other hand, is determined
by the total number of points along each axis, denoted $x_{tot}$,
$y_{tot}$ etc.

Geometrically, the simplest discretization scheme is based on a cubic
grid. This puts the Hamiltonian in the form
\begin{eqnarray}
 & H_{cubic}= & \frac{a}{2}\sum_{k=1}^{xtot}\sum_{l=1}^{ytot}\sum_{m=1}^{ztot}\left[\pi_{klm}^{2}+\left(\varphi_{\left(k+1\right)lm}-\varphi_{klm}\right)^{2}\right.\nonumber \\
 &  & \hspace{-1cm}\left.+\left(\varphi_{k\left(l+1\right)m}-\varphi_{klm}\right)^{2}+\left(\varphi_{kl\left(m+1\right)}-\varphi_{klm}\right)^{2}\right],
\end{eqnarray}
where $\pi$ denotes the conjugate momentum. Following Srednicki \cite{S93},
we identify this as the Hamiltonian of an array of $N=x_{tot}\cdot y_{tot}\cdot z_{tot}$
coupled harmonic oscillators: if we replace the indices $klm$ by
a single $j=1,\ldots N$, 
\begin{equation}
H_{cubic}=\frac{a}{2}\left[\sum_{j=1}^{N}\pi_{j}^{2}+\sum_{i,j=1}^{N}\varphi_{i}K_{ij}\varphi_{j}\right]
\end{equation}
for some coupling matrix $K$. Let $U$ be the matrix that makes $K_{D}\equiv UKU^{T}$
diagonal, and $\Omega=U^{T}K_{D}^{1/2}U$. Treating the list of values
$\varphi\equiv\left(\varphi_{1},\ldots,\varphi_{N}\right)$ as a vector
allows us to write the ground-state density operator of such a system
simply in terms of matrix products,
\begin{equation}
\rho_{0}\left(\varphi,\varphi^{\prime}\right)\propto\exp\left(-\frac{1}{2}\varphi^{T}\Omega\varphi-\frac{1}{2}\varphi^{\prime T}\Omega\varphi^{\prime}\right).\label{eq:groud state}
\end{equation}
Since the numbering of the sites by the index $j$ is arbitrary, we
can choose an ordering such that the partial trace is taken over the
degrees of freedom $\check{\varphi}=\left(\varphi_{n+1},\ldots,\varphi_{N}\right)$.
The reduced density matrix over the remaining $n$ oscillators, $\tilde{\varphi}=\left(\varphi_{1},\ldots,\varphi_{n}\right)$,
is 
\begin{equation}
\rho_{red}\left(\tilde{\varphi};\tilde{\varphi}^{\prime}\right)\equiv\int\left[\prod_{j=n+1}^{N}d\varphi_{j}\right]\rho_{0}\left(\tilde{\varphi},\check{\varphi};\tilde{\varphi}^{\prime},\check{\varphi}\right).\label{eq:integral for rhored}
\end{equation}
Given the form of $\rho_{0}\left(\varphi,\varphi^{\prime}\right)$
in (\ref{eq:groud state}), this is an n-dimensional Gaussian integral.
We decompose

\begin{equation}
\Omega_{N\times N}=\left(\begin{array}{cc}
C_{n\times n} & B_{n\times\left(N-n\right)}^{T}\\
B_{\left(N-n\right)\times n} & A_{\left(N-n\right)\times\left(N-n\right)}
\end{array}\right),
\end{equation}
and define the $n\times n$ matrices 
\begin{equation}
\beta=\frac{1}{2}B^{T}A^{-1}B\;\;\;\;\;\gamma=C-\beta.
\end{equation}
In terms of them, the integration in (\ref{eq:integral for rhored})
gives 
\begin{equation}
\rho_{red}\left(\tilde{\varphi};\tilde{\varphi}^{\prime}\right)\propto\exp\left(-\frac{1}{2}\left[\tilde{\varphi}^{T}\gamma\tilde{\varphi}+\tilde{\varphi}^{\prime T}\gamma\tilde{\varphi}^{\prime}\right]+\tilde{\varphi}^{T}\beta\tilde{\varphi}^{\prime}\right).
\end{equation}
It can be further simplified with the following substitutions: let
$V$ be the matrix that makes $\gamma_{D}\equiv V\gamma V^{T}$ diagonal
and define 
\begin{equation}
\beta^{\prime}\equiv\gamma_{D}^{-1/2}V\beta V^{T}\gamma_{D}^{-1/2}.
\end{equation}
Then let $\beta_{j}^{\prime}$ be the eigenvalues of $\beta^{\prime}$,
let $W$ be the matrix that makes $\beta_{D}^{\prime}\equiv W\beta^{\prime}W^{T}$
diagonal and define the new vector of variables
\begin{equation}
\bar{\varphi}\equiv W\gamma_{D}^{1/2}V\tilde{\varphi}.
\end{equation}
In terms of them, $\rho_{red}$ takes the form
\begin{equation}
{\displaystyle \rho_{red}\left(\bar{\varphi},\bar{\varphi}^{\prime}\right)\propto\prod_{j=1}^{n}\exp\left[-\frac{1}{2}\left(\bar{\varphi}_{j}^{2}+\bar{\varphi}_{j}{}^{\prime2}\right)+\bar{\varphi}_{j}\beta_{j}^{\prime}\bar{\varphi}_{j}^{\prime}\right]}.
\end{equation}
Each term in this product is simply a thermal density matrix for a
simple harmonic oscillator $\bar{\varphi}_{j}$, whose frequency $\omega_{j}$
and temperature $T_{j}$ are related to $\beta_{j}^{\prime}$ by
\begin{equation}
\xi_{j}\equiv\exp\left(-\frac{\omega_{j}}{T_{j}}\right)=\frac{\beta_{j}^{\prime}}{1+\sqrt{1-\left(\beta_{j}^{\prime}\right)^{2}}}.
\end{equation}
Thus it has entropy
\begin{equation}
S_{j}=-\ln\left(1-\xi_{j}\right)-\frac{\xi_{j}}{1-\xi_{j}}\ln\xi_{j}.
\end{equation}
The total entropy of the reduced state, that is, the entanglement
entropy, is simply the sum of these terms.

\section{Tests in flat space-time}

\label{sec:flat} 

The cubic discretization scheme allows us to choose for each site
independently whether the value of the field at that point is to be
traced out or not, thus giving us the freedom to study the entanglement
entropy of regions with arbitrary shape. This reveals an area law,
\begin{equation}
S_{cubic}=0.064\frac{A}{4a^{2}},\label{eq:kappa cubic}
\end{equation}
not only for cubes and parallelepipeds, but also hollow shells (in
which case both the inner and outer surfaces contribute) and in fact
arbitrary configurations. One can, for example, choose a set of unit
cubes such that they approximate a sphere, and still (\ref{eq:kappa cubic})
holds, with the same coefficient. This extends to more convoluted
shapes, even when one portion of the surface comes to within one lattice
site of another. The fact that there are no long-range effects --
and consequently no influence of the global geometry of $\Omega$
on $S$ --, but only the linear scaling with the area, is a useful
stepping stone towards our main question: how the geometry of the
background space-time might affect $S$.

Before we proceed to the Einstein universe, it is also useful to introduce
and test a second discretization scheme. In a spherically symmetric
setting, it is convenient to decompose $\varphi\left(\vec{x}\right)$
in partial waves, $\varphi_{lm}\left(r\right)$. Their angular dependence
is given by the real spherical harmonics $Z_{lm}\left(\theta,\phi\right)$,
so that they are functions only of the radial coordinate. Given the
symmetry of the system, the partial waves are not coupled to each
other, and their contributions to entanglement entropy can be computed
separately: for each $lm$, the radial coordinate is discretized,
and the problem again reduces to an array of coupled harmonic oscillators.
By the same procedure outlined above, we find the contributions $S_{lm}$,
which are summed to obtain the total entanglement entropy $S$. In
practice, the convergence of the sum over $l$ allows us to extrapolate
the exact value of $S$ after computing only a finite number of terms. 

This computation again leads to an area law:
\begin{equation}
S_{sphere}=0.088\frac{A}{4a^{2}}.\label{eq: S sphere flat}
\end{equation}
At first glance, this seems incompatible with result~(\ref{eq:kappa cubic})
by a factor of $\sim3/2$. However, consider a comparison of the entropy
each method yields for a given $\Omega$, namely a sphere of radius
$R$: using the spherical discretization scheme, we find simply $0.088\cdot4\pi R^{2}/4$.
When this $\Omega$ is decomposed into unit cubes, however, the surface
of the actual region is not smooth, and can be shown to have an area
of $6\pi R^{2}$. Thus, the entropy becomes $0.064\cdot6\pi R^{2}/4$,
which agrees reasonably well with the previous result.

Besides the geometry of the traced-out region $\Omega$, we also explore
how the entropy depends on the total space in which the field lives,
$\bar{\Omega}$ (the discrete array, in our model). We argue in the
introduction that $S$ can not depend on geometric properties of the
exterior, such as the infrared cutoff or the shape of $\bar{\Omega}$.
However, if $\bar{\Omega}$ were to become smaller than $\Omega$,
so that all degrees of freedom of the field are traced out, then clearly
the entropy must go to zero. Our simulation allows us to explore this
transition, revealing that the entropy remains constant until $\bar{\Omega}$
comes to within $3a$ of $\Omega$, in spherical symmetry, and as
little as a single lattice site in the cubic array. Again, there is
no evidence of long-range effects in the generation of entanglement
entropy.


\section{Entropy in the Einstein universe}

\label{sec:Einstein} 

Now let us replace the Minkowski space-time by an Einstein universe,
which has the spatial geometry of a 3-sphere. The corresponding line
element can be cast as 
\begin{equation}
ds^{2}=dt^{2}-R_{0}^{2}\left(d\chi^{2}+\sin^{2}\chi\left(d\theta^{2}+\sin^{2}\theta d\phi^{2}\right)\right),\label{eq:Einstein}
\end{equation}
 with $0\leq\chi,\theta\leq\pi$, $0\leq\phi<2\pi$. Because the Einstein
universe is compact, the curvature radius $R_{0}$ also defines a
natural infrared cutoff scale, $L\equiv\pi R_{0}$. At fixed time
$t$, the surface of constant $\chi$ is the boundary of a spherical
region $\Omega$, whose radius is $r=R_{0}\chi$. Its surface area
is $A=4\pi R_{0}^{2}\sin^{2}\chi$, which is upper bounded by $A_{max}=4L^{2}/\pi$
(at $\chi=\pi/2$). When $\chi\ll1$, the local geometry of Eq.~(\ref{eq:Einstein})
approaches that of Minkowski space-time.

The discretization scheme for spherical symmetry outlined above (see
section~\ref{sec:flat}) carries over naturally to this case: the
radial direction is divided into $N=L/a$ discrete points with spacing
$a$, indexed by $j$. For each partial wave $lm$, this leads to
a coupling matrix $K_{lm}$, from which one can calculate the contributions
$S_{lm}$, and consequently $S$.

For spheres that are small compared to the curvature radius, we recover
the flat space area law (\ref{eq: S sphere flat}) to good precision
(considering the limitations of numerical simulations): 
\begin{equation}
S_{curv}=0.092\frac{A}{4a^{2}}.\label{eq:S curv}
\end{equation}
In fact, it can be shown analytically that, in the low-curvature limit,
the coupling matrix reduces to its flat-space counterpart. We also
note that $S$ is the same for $r$ and $\pi R_{0}-r$ -- as expected,
since both generate boundaries with the same area.

The question, then, is how $S$ varies as a function of $A$ if the
sphere is large enough that effects of curvature become noticeable,
i.e., in the limit $r\rightarrow R_{0}/2$, or 
\begin{equation}
\frac{A}{4a^{2}}\rightarrow\frac{A_{max}}{4a^{2}}=\frac{\pi R_{0}^{2}}{a^{2}}.
\end{equation}

\begin{figure}[h!]
\includegraphics[width=9cm]{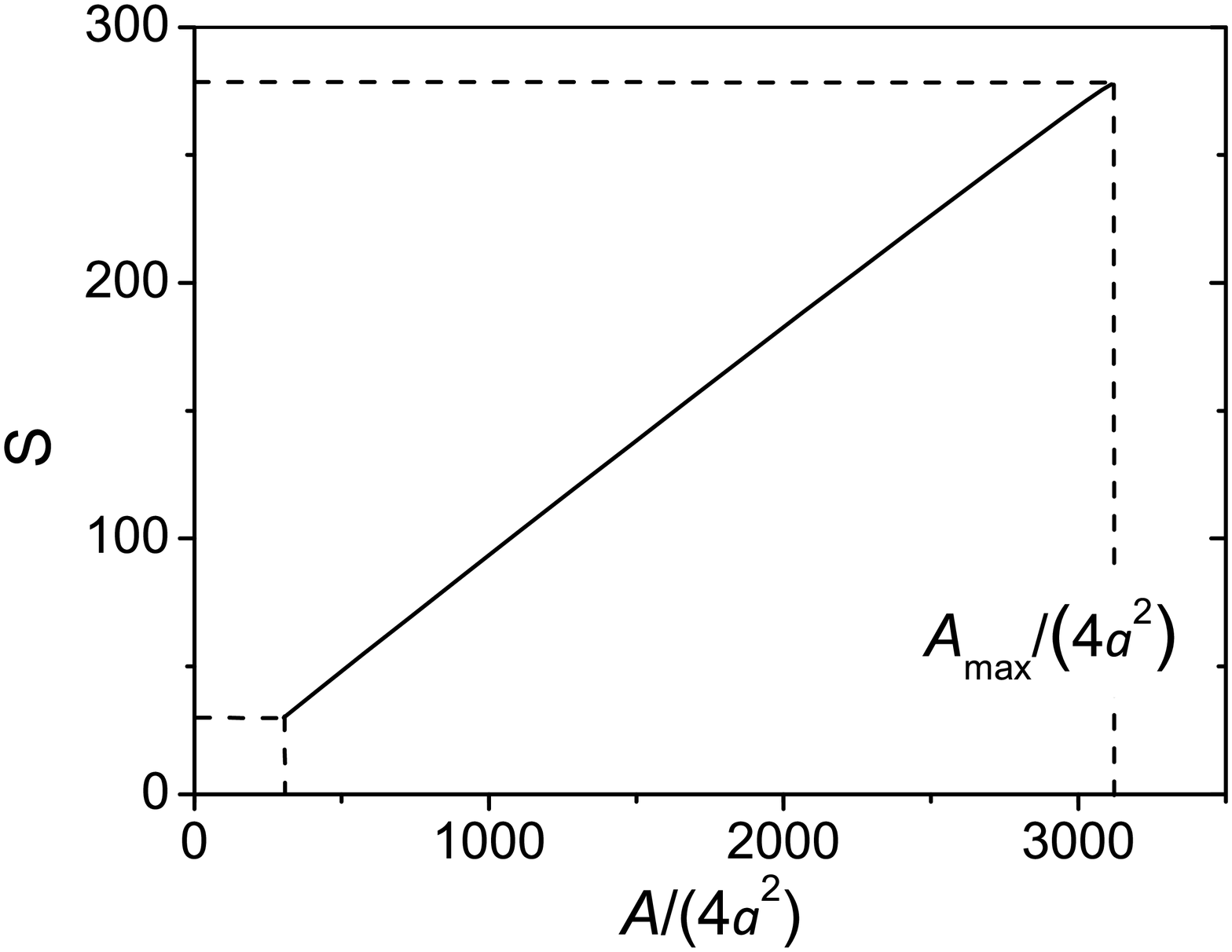} \caption{Entanglement entropy $S$ as a function of surface area $A$ (in units
of $4a^{2}$) for spheres in an Einstein universe ($N=99$). Generally,
curvature effects become non-negligible as $A$ approaches its upper
bound, but the area law for the entropy is unaffected. (The lower
bound $A/\left(4a^{2}\right)\gg1$ is imposed by the semi-classical
approximation; for small areas, the quantum nature of the boundary
itself become relevant.)}

\label{figura1} 
\end{figure}

Fig.~\ref{figura1} gives a clear answer: the area law (\ref{eq:S curv})
holds regardless of the large-scale curvature of the background. 

\section{Conclusions}

\label{sec:finalremaks} 

The entanglement entropy $S$ of a massless scalar field across a
spherical surface in an Einstein universe does not depend on the radius
of the universe, $R_{0}$. In the flat-space limit - a Minkowski universe,
simulated out to a finite radius $R$ - it is not surprising that
$S$ does not depend on the infrared cutoff, because $R$ characterizes
only the region outside the sphere, whereas the entropy can only depend
on shared properties of the interior and the exterior. In curved space,
on the other hand, the curvature radius $R_{0}$ is a meaningful physical
parameter of the entire space, and therefore it is non-trivial that
it does not affect $S$. We understand this independence as an extension
of a trend observed in flat-space studies of the entropy: entanglement
entropy is dominated by short-range interactions, and therefore ``blind''
to large-scale features, including curved background. 

The relevant scale is given by the ultraviolet cutoff of the theory,
$a$. In our model, it is introduced ad hoc, by discretizing the field,
but how it arises from natural laws is a topic of active research.
The cutoff is expected to be of the order of the Planck length, both
for dimensional reasons and to ensure that the area law for entanglement
entropy is compatible with other expressions for the entropy of black
holes. If the curvature of space-time becomes noticeable on that scale,
which characterizes the regime of quantum gravity, the area law is
expected to break down. 

Corrections to the area law also arise if one considers flat space,
but non-zero spin or excited states \cite{GK96,DSS08}, suggesting
that the entropy is no longer determined solely by short-range interactions.
It would be interesting to determine how the entropy responds to background
curvature in these cases. 

\acknowledgments 

I am grateful to George Matsas for his supervision and guidance during
the work that led to this paper. The research was supported by scholarships
from Conselho Nacional de Desenvolvimento Cient\'{i}fico e Tecnol\'{o}gico (CNPq,
process number 136114/2009-1) and Funda\c{c}\~{a}o de Amparo \`{a} Pesquisa do
Estado de S\~{a}o Paulo (Fapesp, process number 2009/10774-8), and completed
at Perimeter Institute for Theoretical Physics. Research at Perimeter
Institute is supported by the Government of Canada through Industry
Canada and by the Province of Ontario through the Ministry of Research
and Innovation.

\end{document}